\documentclass[12pt]{iopart}
\usepackage{graphics}
\usepackage{graphicx}
\usepackage{amssymb}

\begin{document}
\title{Heavy Quarkonia Production in p+p Collisions from the PHENIX Experiment}
\author{Abigail Bickley for the PHENIX\footnote{For the full list of PHENIX authors and acknowledgements, see Appendix 'Collaborations' of this volume} Collaboration}
\address{Department of Physics, University of Colorado, Boulder, Colorado, 80309-0390, USA} 
\ead{bickleya@colorado.edu}

\begin{abstract}
Quarkonia provide a sensitive probe of the properties of the hot dense medium
created in high energy heavy ion collisions.  Hard scattering processes result
in the production of heavy quark pairs that interact with the collision medium
during hadronization.  These in-medium interactions convey information about
the fundamental properties of the medium itself and can be used to examine the
modification of the QCD confining potential in the collision environment.
Baseline measurements from p+p and d+Au collision systems are used to distinguish
cold nuclear matter effects while measurements from heavy ion collision systems
are used to quantify in-medium effects.  The PHENIX experiment has the capability of
detecting heavy quarkonia at $1.2<|\eta|<2.2$ via the $\mu^+\mu^-$
decay channel and at $|\eta|<0.35$ via the $e^+e^-$ decay channel.
Recent runs have resulted in the collection of high statistics p+p data sets that provide
an essential baseline reference for heavy ion measurements and allow for further
critical evaluation of heavy quarkonia production mechanisms.  The latest PHENIX results for the
production of the $J/\psi$ in p+p collisions are presented and future prospects for $\psi'$, $\chi_{c}$ 
and $\Upsilon$ measurements are discussed.  

\end{abstract}

%\maketitle

\section{Introduction}
The study of quarkonia in p+p collisions provides a useful tool for probing heavy quark production mechanisms.  Heavy quarks are predominantly generated in hadronic collisions via hard processes involving gluonic diagrams, but the details of hadronization remain unclear.  Several theoretical models, with varying degrees of success, have been proposed to explain how a heavy quark pair evolves to form a bound quarkonium state.  The color-singlet model generates $J/\psi$ particles from color singlet $c\bar{c}$ pairs that are in the same quantum state as the final $J/\psi$ \cite{CSM}.  However, the $J/\psi$ production cross section is underestimated by an order of magnitude \cite{CDF}.  The NRQCD color-octet model \cite{COM} includes color-octet state $c\bar{c}$ pairs in addition to color-singlet state pairs that radiate soft gluons during hadronization to form a $J/\psi$.  Contrary to expectation, the color-octet matrix elements, derived from experimental data, are found to be non-universal \cite{COM_matrix}.  Additionally, a large transverse polarization is predicted at high $p_{T}$ that is not observed in the data.  
The color evaporation model provides a phenomenological approach to hadronization \cite{CEM}.  The charmonium states are formed in proportions determined by experiment for any $c\bar{c}$ pair below the $D\bar{D}$ threshold and hadronization occurs through the emission of soft gluons.  
To date, the most successful model of quarkonia production is a perturbative QCD approach involving 3-gluon mechanisms which is able to reproduce both the experimentally observed $J/\psi$ cross section and polarization \cite{Khoze}.  Within this model, the hadronization mechanism includes a channel involving the fusion of a symmetric color-octet state with an additional gluon.  The measurements reported in this paper provide further tests for these production mechanisms.

\section{PHENIX Experiment}
The PHENIX experiment is designed to detect heavy quarkonia at forward rapidity ($1.2<|\eta|<2.2$) via the $\mu^+\mu^-$ decay channel and at mid-rapidity ($|\eta|<0.35$) via the $e^+e^-$ decay channel \cite{NIM}.  At forward rapidity the muon detectors are composed of cathode strip tracking chambers in a magnetic field and alternating layers of steel absorber and Iarocci tube planes that allow muons to be tracked and reconstructed over an acceptance of $\Delta\phi=360^{\circ}$.  At mid-rapidity the central arm tracking detectors, composed of drift chambers, ring imaging \v{C}erenkov detectors, and electromagnetic calorimeters, are used to detect electrons in two arms each covering $\Delta\phi=90^{\circ}$ in azimuth.  During the 2005 run of the Relativistic Heavy Ion Collider the PHENIX experiment sampled $3.8~pb^{-1}$ of p+p collisions.  From this data set a total of 8000 (1500) $J/\psi$ have been reconstructed in the dimuon (dielectron) channel \cite{ppg069}.

\section{Transverse Momentum Dependence of the $J/\psi$ Cross section}
\begin{figure}
\begin{center}
$\begin{array}{c@{\hspace{0.25in}}c}
\includegraphics[width=2.8in]{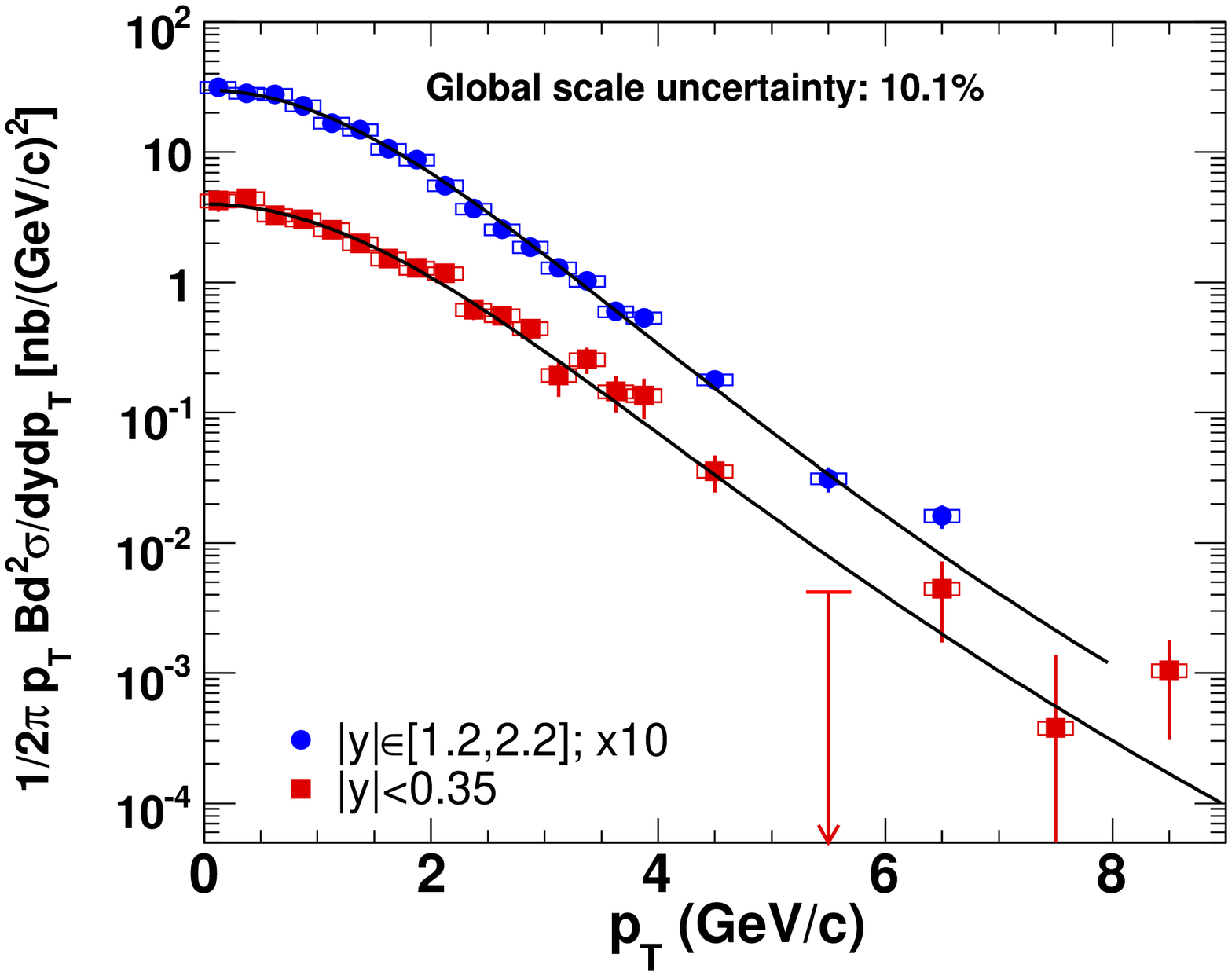}
\includegraphics[width=2.8in]{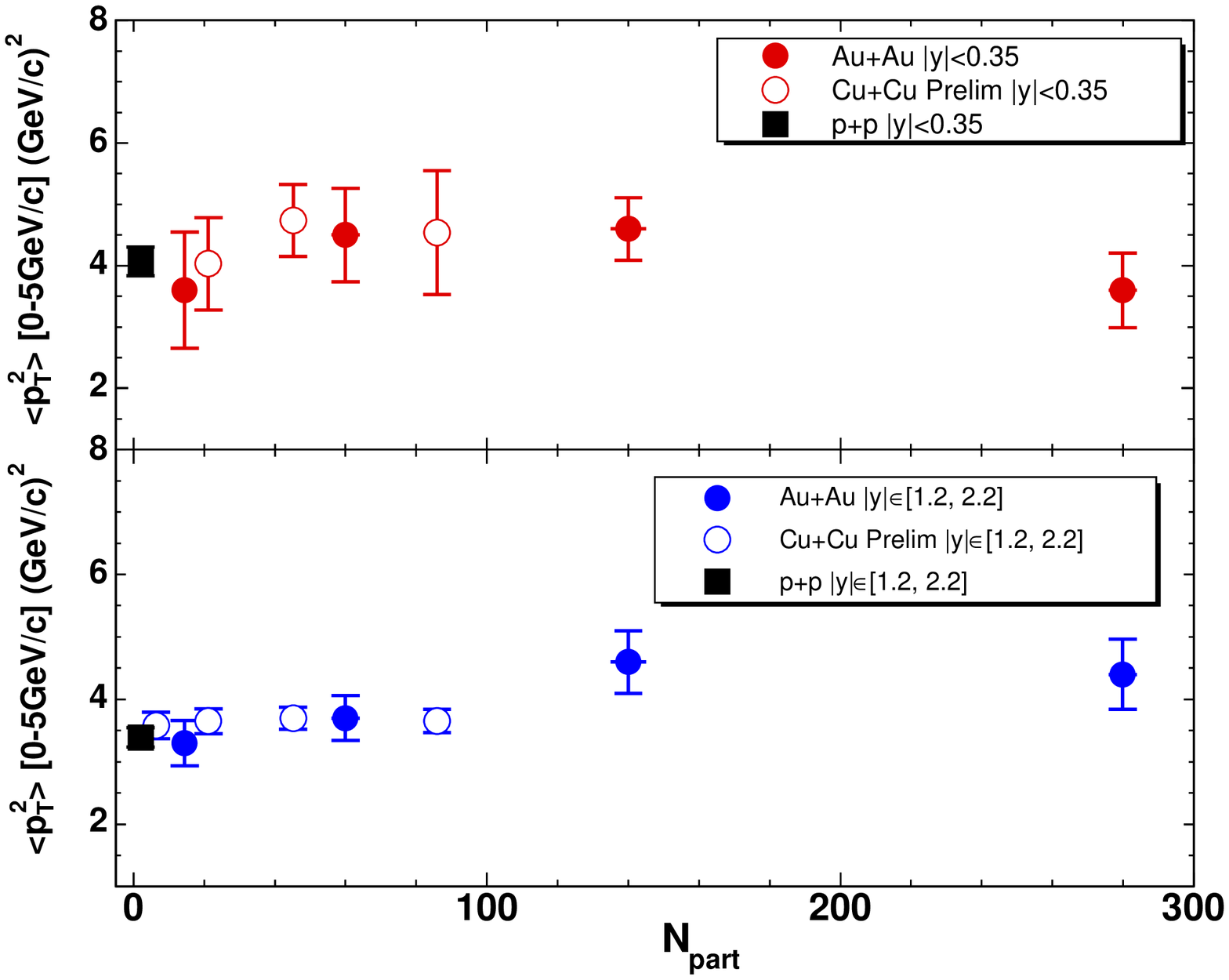}
\end{array}$
\end{center}
\caption[jpsi_pt]{Left: The $J/\psi$ differential cross section times di-lepton branching ratio versus $p_T$ at mid and forward rapidity.  The vertical error bars are the statistical and point-to-point uncorrelated error and the boxes are the point-to-point correlated systematic error. The solid lines are the fits \cite{ppg069}.  Right:  The $J/\psi$ $\langle p_T^2 \rangle$ truncated at 5 GeV/$c$ plotted versus $N_{part}$ in p+p, Cu+Cu and Au+Au collisions \cite{ppg069, ppg068, QM05_hugo}.}
\label{jpsi_pt}
\end{figure}

As shown in Fig.~\ref{jpsi_pt}, the transverse momentum dependence of the $J/\psi$ cross section can be mapped to 9 GeV/$c$ using the new high statistics data.  A comparison of the spectral shapes at mid and forward rapidity reveals that the $p_{T}$ spectrum is softer at forward rapidity.  This is likely a result of the increased longitudinal momentum at forward rapidity, consequently there is less energy available in the transverse direction.  

The spectra are fit with the functional form, \mbox{$A\times(1+(p_T/B)^2)^{-6}$}, to extract the $\langle p_T^2\rangle$.   At mid-rapidity $\langle p_T^2\rangle = 4.14\pm0.18\pm^{0.30}_{0.20}$ (GeV/$c$)$^{2}$ and the $\chi^2$ per degree of freedom ($\chi^{2}/$ndf) is $23/19$.  At forward rapidity $\langle p_T^2\rangle = 3.59\pm0.06\pm0.16$ (GeV/$c$)$^2$ and the $\chi^{2}/$ndf is $28/17$.  If the exponent in the fit function is allowed be a free fit parameter a slightly better fit is achieved at forward rapidity, $\chi^{2}/$ndf = $20/16$, but the $\langle p_T^2\rangle$ is not significantly modified, $\langle p_T^2\rangle = 3.68$ (GeV/$c$)$^2$.  The previous results from the 2003 run \cite{ppg038} yielded a significantly lower $\langle p_T^2\rangle$ at forward rapidity, even though within errors the $p_{T}$ spectra agree.  The increased statistics of the 2005 data allow for an improved understanding of the shape of the $p_T$ spectrum due to the extended range in $p_{T}$ and the finer binning at low $p_{T}$.  The 2003 p+p results have been revisited and it was found that the systematic error was underestimated.  A reanalysis of the 2003 run d+Au data is underway to determine how this effects the interpretation of that data set.  

A comparison of the $\langle p_T^2\rangle$ found in p+p collisions with that found in Au+Au and Cu+Cu collisions as a function of centrality is shown in Fig.~\ref{jpsi_pt} \cite{ppg068, ppg069, QM05_hugo}.  Because the heavy ion $p_{T}$ spectra only extend to 5 GeV/$c$ the calculation of the $\langle p_T^2\rangle$ is truncated at 5 GeV/$c$ for all data sets.  Good agreement is found between the peripheral heavy ion data and the p+p measurements.  

%Figure 1

\section{Rapidity Dependence of the $J/\psi$ Cross section}
\begin{figure}
\begin{center}
$\begin{array}{c@{\hspace{0.25in}}c}
\includegraphics[width=2.8in]{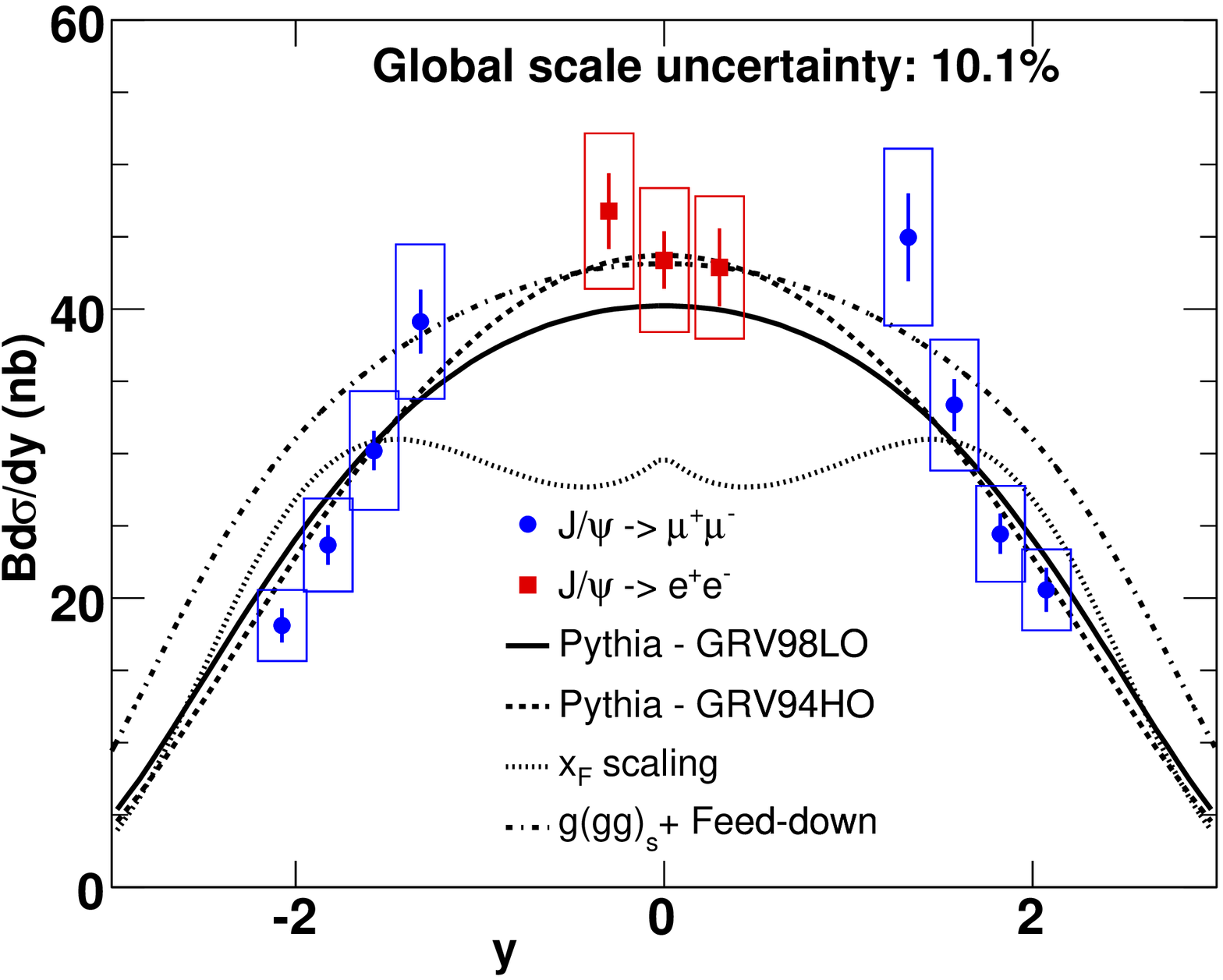}
\includegraphics[width=2.8in]{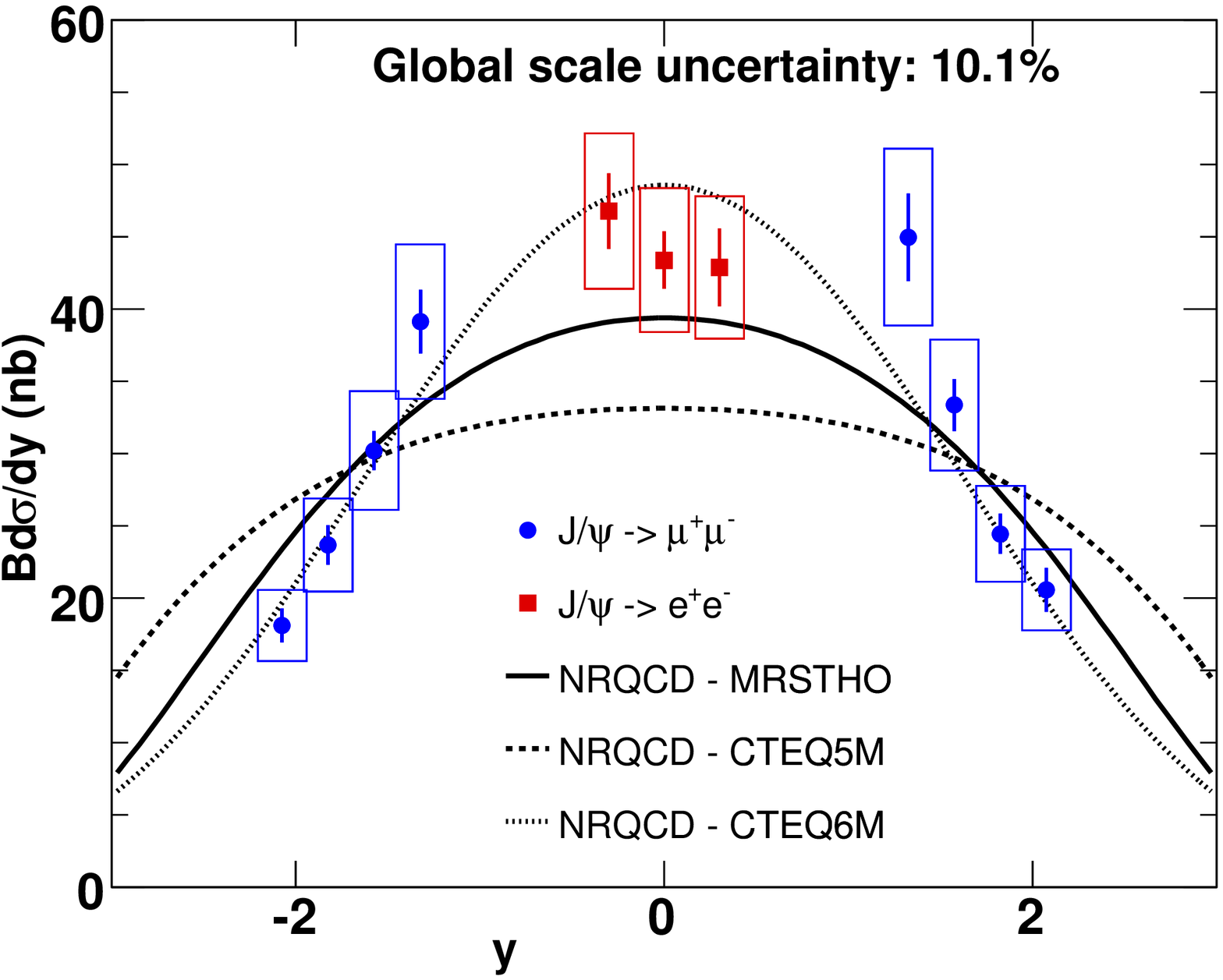}
\end{array}$
\end{center}
\caption[jpsi_y]{$J/\psi$ cross section in p+p collisions as a function of rapidity \cite{ppg069} plotted with various parton distribution functions and model calculations for comparison \cite{Khoze, ramona, bratk}.} 
\label{jpsi_y}
\end{figure}

The rapidity dependence of the $J/\psi$ cross section has been mapped over the range $-2.2<y<2.2$, Fig.~\ref{jpsi_y}.  The data exhibit a slight flattening over the rapidity range $|y|<1.5$.  However, the systematic errors on the mid and forward rapidity points are independent, thus a narrower distribution is not excluded.  A comparison of the shape of the distribution with various parton distribution functions and model calculations \cite{Khoze, ramona, bratk} shows that many of the models, including the perturbative QCD approach involving 3-gluon mechanisms, are inconsistent with the steepness of the slope at forward rapidity and the slight flattening observed at mid-rapidity.

%Figure 2

\section{Other Quarkonia}
In addition to the $J/\psi$, the PHENIX detector has the capability of measuring the $\Upsilon$, $\chi_{c}$ and $\psi'$ quarkonia states in p+p collisions.  Preliminary PHENIX measurements of the $\Upsilon$ as a function of rapidity and collision energy are available \cite{leitch}.  The analysis of the $\chi_{c}$ and $\psi'$ states is still ongoing.  The 2006 p+p run provides a factor of 3 improvement in the statistics available for these measurements and the results will be forthcoming.  

\section{Summary}
With the existing high statistics data the PHENIX experiment has entered an era of precision $J/\psi$ measurements.  The p+p data exhibit several features that provide challenges for production models.  The $p_{T}$ spectrum has been measured over the range $0< p_{T} < 9$ GeV/$c$ and is softer at forward rapidity than at mid-rapidity.  Furthermore, an improved assessment of the $J/\psi$ $\langle p_T^2 \rangle$ is possible with the 2005 data set and good agreement is found between the peripheral heavy ion data and the p+p measurements.  The $J/\psi$ cross section as a function of rapidity is slightly flat and falls off rapidly at forward rapidity.  This behavior is not well reproduced by the available model calculations.  Future PHENIX data will shed light on these processes and open additional exciting avenues of quarkonia measurements.

\end{document}